\documentstyle[11pt]{article}
\newcommand{\be}{\begin{equation}}
\newcommand{\ee}{\end{equation}}
\newcommand{\ba}{\begin{eqnarray}}
\newcommand{\ea}{\end{eqnarray}}

\begin{document}

\title{\vspace{-3.0cm} \hspace{0.0cm} \hspace*{\fill} \\[-5.5ex]
\hspace*{\fill}{\normalsize LAUR-96-1298} \\[1.5ex]
{\huge {\bf Hydrodynamical analysis of
symmetric nucleus-nucleus collisions at CERN/SPS energies}}}

\author{
U. Ornik${}^2$\thanks{E. Mail: ornik@warp.soultek.de}{\ },
M. Pl\"umer${}^3$\thanks{E. Mail: pluemer@Mailer.Uni-Marburg.DE}{\ }, 
B.R. Schlei${}^1$\thanks{E. Mail: schlei@t2.LANL.gov}{\ },  
D. Strottman${}^1$\thanks{E. Mail: dds@LANL.gov}{\ } and
R.M. Weiner${}^3$\thanks{E. Mail: weiner@Mailer.Uni-Marburg.DE}\\[1.5ex]
{\it ${}^1$Theoretical Division, Los Alamos National Laboratory, 
Los Alamos, NM 87545}\\
{\it ${}^2$SoulTek Internet Services, Software Center 5, Marburg, Germany}\\
{\it ${}^3$Physics Department, Univ. of Marburg, Marburg, Germany}
}

\date{April 11, 1996}

\maketitle

\vspace*{-1.0cm}
\begin{abstract}
We present a coherent theoretical study of ultrarelativistic heavy-ion data
obtained at the CERN/SPS by the NA35/NA49 Collaborations using
3+1-dimensional relativistic hydrodynamics. We find excellent agreement with
the rapidity spectra of negative hadrons and protons and with the
correlation measurements in two experiments:
$S+S$ at 200 $AGeV$ and $Pb+Pb$ at 160 $AGeV$ (preliminary results). Within
our model this implies that for $Pb+Pb$ ($S+S$) a quark-gluon-plasma  of
initial volume 174 $fm^3$ (24 $fm^3$) with a lifetime 3.4 $fm/c$  (1.5
$fm/c$) was formed. It is found that the Bose-Einstein correlation
measurements do not determine the maximal effective radii of the hadron
sources because of the large contributions from resonance decay at small 
momenta. Also within this study we present an NA49
acceptance corrected two-pion Bose-Einstein correlation function in the
invariant variable, $Q_{inv}$.\\
\end{abstract}

\hbadness=10000

\vspace{-0.5cm}

\newpage

\section{Introduction}

In the last decade many experiments have been performed in the
attempt to find evidence for the existence of a quark-gluon plasma (QGP).
Ever higher energies and/or masses have been involved in order to increase
the lifetime of the system either by increasing the initial energy density
or the size of the system.
The probability of preparing a strongly interacting system that
shows thermodynamical behaviour and therefore is treatable by well known
thermodynamical or fluid dynamical methods increases with the size of the
system.
In this paper we describe a comprehensive hydrodynamical study of
data taken at the CERN/SPS by the NA35 and the NA49 Collaborations.
Our investigation takes into account all available data spectra and
correlation functions for different particle species. We present
results of the analysis \cite{bernd7} of $Pb+Pb$ at 160 $AGeV$ using
relativistic fluid dynamics and assuming an equation-of-state
containing a phase transition. These results are discussed by
comparison to our previous findings for $S+S$ at 200 $AGeV$ which were
obtained by applying the same computer code HYLANDER \cite{udo}.

Many hydrodynamical models \cite{hydro1,hydro2,hydro3} are available 
which describe the dynamics of relativistic heavy-ion collisions.
HYLANDER belongs to the class of models which apply 3+1-dimensional
relativistic one-fluid-dynamics. It provides fully three-dimensional 
solutions of the hydrodynamical relativistic Euler-equations 
\cite{euler}. 
HYLANDER has been successfully applied at SPS energies to the reaction
$Au+O$ and especially to $S+S$, a reaction to which we refer several 
times in this paper. Here HYLANDER was used to
reproduce \cite{jan} simultaneously mesonic and baryonic rapidity
and transverse momentum spectra of the $S+S$ reaction at 200 $AGeV$. 
Corresponding measurements had been performed by the
NA35 Collaboration \cite{wenig}. Based on the successful description of
the measured single-inclusive spectra,
predictions were made for Bose-Einstein correlation (BEC)
functions \cite{bernd2,bernd3}. Those predictions agree
quantitatively with the measurements \cite{alber,QM95} ({\it cf.}
also Fig. 5a). The model also reproduces the photon data for $S+Au$
collisions at SPS energies \cite{axel} and gives a simple explanation
for the ``soft-$p_{\perp }$ puzzle'' \cite{udo91} and the complex
behaviour of the radii extracted from pion and kaon correlations and
explains the difference in the extracted radii for pions and kaons
in terms of a cloud of pions due to the decay of resonances which
surrounds the fireball (pion halo) \cite{bernd2,csorgo}.

In the present paper we exhibit the space-time geometries of the
real hadron sources, i.e., the freeze-out hypersurfaces. Results for
$S+S$ at 200 $AGeV$ and $Pb+Pb$ at 160 $AGeV$ are directly compared.
For the first time we show our results of effective radii compared
to the BEC data of the NA35 Collaboration. Furthermore, we present a
calculation of the detector-acceptance-corrected two-pion
correlation function $\tilde{C}_2(Q_{inv})$ in the invariant
variable, $Q_{inv}$. In doing
so we would like to bring the readers attention also to refs.
\cite{bernd4,bernd5} where all the features of these specific types of 
two-particle BEC functions were extensively discussed. The
following discussion is a self-consistent description of two
different heavy-ion experiments.  Here we reproduce simultaneously
single inclusive spectra of negative  hadrons and protons and
Bose-Einstein correlations of identical pions using only one and the
same equation of state. The contents of this paper represents a
summary of many contributions
\cite{bernd7,udo,jan,bernd2,bernd3,bernd1} in an effort to give a
description of heavy ion data taken at the CERN/SPS in terms of 3+1
dimensional relativistic hydrodynamics.\\

\section{Modeling the hydrodynamical solutions}
HYLANDER can use initial conditions ranging between
the extremes defined by the Landau \cite{landau} and the
Bjorken \cite{bjorken} initial conditions. One has to specify an
equation of state and a set of parameters which describe the initial
conditions. In view of our former results for $S+S$ at 200 $AGeV$,
the same equation of state (EOS) is used for the treatment of
$Pb+Pb$ at 160 $AGeV$. It exhibits a phase transition to a quark-gluon
plasma at a critical temperature $T_C=200\:MeV$ ({\it cf.} refs. 
\cite{redlich,phdudo,dipljan}). This EOS, which has no 
dependence on the baryon density, is plotted in Fig. 1.
In the following we sketch the basic features of the model which was
introduced in ref. \cite{jan}. Our model uses the five parameters,
$K_L$, $\Delta$, $y_\Delta$, $y_m$ and $\sigma$, which are explained
below.\\

A reaction of two baryonic fluids leads to a deceleration
of the projectile and target baryonic currents and thus to the spread of
their width in momentum space. For the initial baryon density
distribution in rapidity we write 
\begin{equation}
\frac{db}{dy}\:=\:C_y\:\left[e^{-(y-y_m)/2\sigma^2}\:+
\:e^{-(y+y_m)/2\sigma^2}\right]\:, 
\label{eq:dbdy} 
\end{equation}

where $\pm y_m$ give the positions in rapidity of the two maxima and
$\sigma$ gives the width of the baryonic density  distribution after
the collision, respectively. $C_y$ is a normalization constant, the
value of which can be determined from the requirement that when eq.
(\ref{eq:dbdy}) is integrated over the whole accessible rapidity interval, 
it is equal to the total baryon number of the system. A spatial baryon
distribution ({\it cf.} Fig. 2a) is then easily  derived by
evaluating
\begin{equation}
B_0(z)\:=\:\frac{1}{\pi R^2}\frac{db}{dy}\frac{dy}{dz}\:,
\label{eq:b0z}
\end{equation}

where $R$ is the mean radius of the initially radially smeared out
cylindrical fireball ({\it cf.} ref. \cite{bernd3}). In our model we
impose an initial rapidity field $y(z)$ on the fluid. Its modulus is
presumed to be a function only of the longitudinal coordinate $z$
with its shape constrained by two boundary values: The rapidity
should vanish at $z=0$ and asymptotically reach its maximum value
$y_{cm}$ at $z=\pm t_a$ ($y_{cm}$ being the center of mass
rapidity and $t_a$ the time it takes for two nuclei at the speed of
light to penetrate each other, respectively). We parameterize the
function $y(z)$ by the slope parameter $a_y$ as
\begin{equation}
y(z)\:=\:y_{cm}\:\tanh \left[ a_y\:|z| \right]\:.
\label{eq:yofz}
\end{equation}

Rather than using $a_y$ as a free parameter in our model, we use the
two parameters $\Delta$ and $y_\Delta$ determining the slope
parameter $a_y$ in eqn. (\ref{eq:yofz}) through
\begin{equation}
y(z=\pm \Delta/2)\:=\:y_\Delta\:.
\label{eq:ydel}
\end{equation}

In eq. (\ref{eq:ydel}) the quantity $\Delta$ is the spatial 
longitudinal extension
of the initial fireball (in ref. \cite{jan} called the ``Landau
volume'') and $y_\Delta$ is the absolute value of rapidity at
$z=\pm\Delta/2$.\\

The kinetic energy of the two incoming baryonic fluids
is converted into internal excitation (thermal
energy) of a third fluid which is created in the central region. The
relative fraction $K_L$ of the thermal energy inside the initial
fireball volume is another free parameter which fixes the initial
state of the formed fireball.\\

In table 1 we show the choice of the five parameters which leads to the
reproduction of the experimental $S+S$ data taken by the NA35
Collaboration. Additionally, it was there assumed that due to
experimental uncertainty for the centrality of the collision, only
85$\%$ of the total available energy and the total baryon number
have been observed. In Fig. 2a the initial longitudinal
distributions of energy density, as well as the rapidity, normalized
to their maximum values are plotted against the longitudinal
coordinate $z$ (since we deal with a symmetric system, only the
distributions for $z\geq 0$ are shown). Due to the
constraints of energy conservation the choice of the initial
parameters $\Delta$, $y_\Delta$ and $\sigma$ leads to a limited 
two-parameter space for the variables $K_L$ and $y_m$, respectively. 
The limited two-parameter space is shown for $S+S$ in Fig. 2b. 
The crossed lines indicate the choice of the parameter pair for the 
reproduction of the heavy-ion data.\\

For $Pb+Pb$ at 160 $AGeV$ we reduce our five-parameter space to
a two-parameter space based on the results we obtained for $S+S$
at 200 $AGeV$. We assume that the parameter for 
the longitudinal extension of the central fireball scales with $2 R /
\gamma$  compared to the parameter for $S+S$ at 200 $AGeV$
\cite{bernd7}; $\gamma$ is the Lorentz contraction factor. By
coincidence there is a factor two increase in the longitudinal
extension, $\Delta$, going from the reaction  $S+S$ at
200 $AGeV$ to $Pb+Pb$ at 160 $AGeV$. Also the time, $t_a$, it takes
for two nuclei at the speed of light to penetrate each other is
increased by a factor two. For $S+S$ the initial longitudinal
velocity field $v_\parallel(z)=\tanh(y(z))$ increases as a function
of $z$ within the initial fireball almost linearly as in the Bjorken
initial condition scenario (there $v_{Bjorken}(z)=z/\tau$; $\tau$ is
a time scale which can be identified with $t_a$). Since $\Delta$ and
$t_a$ each scale with a factor two, by using the Bjorken scaling argument
for the initial longitudinal velocity field, the increase of the initial
longitudinal rapidity field with the coordinate $z$ is reduced almost
by a factor $0.5$ for $Pb+Pb$ compared to $S+S$ which results in the
same choice for the initial parameter $y_\Delta=y(z=\Delta/2)$ for
both reactions.\\

We have checked that the variation of the width $\sigma$ of the
initial baryon density distribution to a large extent does not affect
the calculation of rapidity spectra. Therefore, the value of $\sigma$
remained unchanged. The values of $\Delta$, $y_\Delta$ and $\sigma$
are given in table 1. Thus we are left with {\it only two}
parameters: the relative fraction of the thermal energy
$K_L$ in the central fireball and the rapidity $y_m$ at the maximum
of the initial baryon distribution. Fig. 2b shows the corresponding
limited two-parameter space for $Pb+Pb$ at 160 $AGeV$ due to energy
conservation. The crossed lines indicate our choice for the two
parameters which are determined by fitting the rapidity spectrum of
negative hadrons of the preliminary NA49 data. All other spectra and
correlation functions, which will be discussed below, are
predictions of the model. Due to the preliminary state of the NA49
data we decided to perform a calculation for the choice of an impact
parameter $b_{imp}=0$. Fig. 2a shows the initial longitudinal
distributions of energy density as well as the rapidity, normalized
to their maximum values ({\it cf.} table 1) as a function of the 
longitudinal coordinate $z$.\\

\section{Discussion of the hydrodynamical solutions}

Once the initial conditions and the equation of state are specified,
one obtains an unambiguous solution from the hydrodynamical
relativistic Euler-equations. In our calculations we assume that
hadronization occurs for all particle species at the same fixed
freeze-out energy density $\epsilon_f$. Since our equation of state
is not dependent on the baryon density, the freeze-out
energy density easily translates into a fixed freeze-out temperature
$T_f$. Our choice for the freeze-out temperature is $T_f=139 MeV$.\\

The choice of a fixed freeze-out temperature, $T_f$,
determines the final space-time geometry of the hydrodynamically
expanding fireball. In Fig. 3a we have plotted the freeze-out regions
at different times for directly produced hadrons in the $z-r$
plane for $S+S$ at 200 $AGeV$ and for $Pb+Pb$ at 160 $AGeV$,
respectively. Each line represents the freeze-out hypersurface at
fixed times $t$. One can see very easily that the solutions we
obtained from our numerical analysis represent an evolution of
initially disk-shaped fireballs which emit hadrons from the very
beginning of their formation. While the relativistic fluids expand
in longitudinal and in transverse directions, the longitudinal
positions of the freeze-out points increase their distance relative
to the center. Due to the effect of transverse inwardly moving
rarefaction waves the transverse freeze-out positions move towards
the center of the fireball ({\it cf.} also ref.
\cite{bernd1}). In the late stage of the hydrodynamical expansion
the hadron-emitting fireballs separate into two parts while cooling
down until they cease to emit.\\

In Fig. 3b we give the full three-dimensional views of the freeze-out
hypersurfaces of directly emitted hadrons for $S+S$ at 200 $AGeV$ as
well as for $Pb+Pb$ at 160 $AGeV$. The freeze-out space-time geometries
show a similar behaviour for both heavy-ion reactions although the
longitudinal and transverse sizes of the systems differ approximately
by a factor two.\\ 

Above we have argued that the initial parameters $\Delta$ and $y_\Delta$
have geometrical scaling features, while $\sigma$ remains unchanged.
The only parameters which were chosen freely (except for the
constraint of energy conservation) are $K_L$ and $y_m$ ({\it cf.}
Fig. 2b). Consistent with expectation the degree of stopping and
thermalization is higher in $Pb+Pb$ and the amount of thermal energy
(represented by the parameter $K_L$) in the central fireball increases
from 43\% $(S+S)$ to 65\% $(Pb+Pb)$. The location of the maximum density 
of the two baryon currents $y_m$ in rapidity space is significantly shifted 
into the central rapidity region. The baryons for $Pb+Pb$ are almost
stopped. The resulting high baryonic density in the $Pb+Pb$ case of
2.14 $fm^{-3}$ in the center is three times higher than 
in $S+S$. In the $Pb+Pb$ case 73\% of the baryons are initially located 
in the central region compared to  only 49\% in the $S+S$ case ({\it cf.} 
Fig.2a).\\

In Fig. 3c we give for both reactions three-dimensional
views of corresponding transverse velocities at freeze-out. The two
systems each show a maximum value for the transverse velocities: for
$S+S$ at 200 $AGeV$ we obtain a maximum transverse velocity
$u_\perp^{max}(S)=0.43$ whereas for $Pb + Pb$ at 160 $AGeV$ we get
$u_\perp^{max}(Pb)=0.61$. But the slope of the transverse increase of
the transverse velocity $u_\perp$ is smaller for the system $Pb+Pb$
compared to the system $S+S$. The reason for this different behaviour
has its origin in the different choice of the initial parameters:
due to a higher initial thermal energy density the internal
pressure of the system $Pb+Pb$ is increased compared to $S+S$, 
resulting in a relatively faster
longitudinal and transverse expansion of the relativistic fluid. The
rarefaction waves in $Pb+Pb$ also move inwardly faster and inhibit the
creation of an adequate large transverse velocity field at
freeze-out, yielding a smaller transverse slope for $u_\perp$
compared to $S+S$.\\

From our calculations we find that for $Pb+Pb$ a quark-gluon-plasma
with an initial volume of 174 $fm^3$ was formed while for $S+S$
there is a QGP with an initial volume of 24 $fm^3$. The maximum
lifetime of the fireball is increased from 6.9 $fm/c$ $(S+S)$ up to
14.5 $fm/c$ for $Pb+Pb$.  We observe approximately the same behaviour
for the lifetimes of the QGP. Whereas in $S+S$ the lifetime of the
QGP was short (1.5 $fm/c$), in the $Pb+Pb$ case a QGP persists for
3.4 $fm/c$. 
The latent heat of the phase transition is of the order of 1 $GeV/fm^3$
({\it cf.} Fig. 1). Taking into account that the initial volume of the 
$Pb+Pb$ system is approximately eight times larger than in the $S+S$ case, 
we note that the lifetime increases slower than does the volume. 
This effect is due to the cooling by transverse rarefaction waves 
which are only sensitive to the difference in the transverse radius 
which differs approximately by a factor two from $S+S$ to $Pb+Pb$.\\

\section{Single inclusive spectra and Bose-Einstein correlations}

In the following we discuss some of the results for the single- and
double-inclusive spectra of mesons and baryons. All calculations are
based on thermal as well as on chemical equilibrium. In both types of
spectra we include the effect of resonance decays. The influence of
partial coherence \cite{bernd3} will not be considered here.\\

The calculation of single particle inclusive spectra with HYLANDER was
extensively discussed in ref. \cite{jan}.
In particular, the momentum distributions are calculated in terms 
of the generalized Cooper-Frye formula (see ref. \cite{fred}),
where explicitly a baryon and a strangeness chemical potential
have to be taken into account.
These potentials have to be introduced, because the assumption of
chemical equilibrium requires zero strangeness and (in general)
nonzero baryon density at each freeze-out point.
In detail one has to solve the following system of equations for each 
surface point of given baryon density $b$ in its rest frame:
\begin{eqnarray}
&&\sum_i b_i n_i(\mu_B,\mu_S)\:=\:b\:,
\nonumber\\
&&\sum_i s_i n_i(\mu_B,\mu_S)\:=\:s\:=\:0\:.
\label{eq:chemeq}
\end{eqnarray}

The index $i$ enumerates all resonances and their anti-particles.
In eq. (\ref{eq:chemeq}) $b_i$, $s_i$ and $n_i(\mu_B,\mu_S)$ 
denote the corresponding baryon number, strangeness and number density 
of the $i$th resonance, respectively. For $S+S$ at 200 $AGeV$ we
find average values for the baryonic chemical potential $\langle \mu_B
\rangle =284\:MeV$ and for the strangeness chemical potential
$\langle \mu_S\rangle =44\:MeV$ using $T_f=139\:MeV$ for the
freeze-out temperature. For $Pb+Pb$ at 160 $AGeV$ we find average
values $\langle \mu_B\rangle =363\:MeV$ and $\langle \mu_S\rangle =69\:MeV$
using the same freeze-out temperature. As a result, we obtain the
chemical compositions of hadrons at freeze-out listed in table 2
({\it cf.} also ref. \cite{jan}).
We stress that due to our choice of the freeze-out temperature of
$T_f=139\:MeV$, there are large resonance contributions ($40\%-50\%$)
to the pionic spectra.\\

Figs. 4a and 4b show our model calculations for negative hadrons $h^-$
and protons $p$ for $S+S$ at 200 $AGeV$ and $Pb+Pb$ at 160 $AGeV$,
respectively. The spectra of negative hadrons $h^-$ are made up by
contributions from negatively charged pions and kaons (directly
emitted ones plus those originating from the decay of resonances).
The $K^-$ spectra consist mainly of directly emitted kaons; less than
$10\%$ come from the $K^\star$ resonance. In the case of the NA35
data we obtained our fit parameters ({\it cf.} table 1) from a
simultaneous fit to the rapidity and transverse momentum spectra for
the negative hadrons $h^-$ and the protons $p$ (for the $S+S$
reaction the proton spectra do
not include contributions from $\Lambda$ decays \cite{jan} 
 while
they do in the treatment of the reaction $Pb+Pb$). By
fitting the $h^-$ rapidity spectrum of the NA49 data we also made
sure that the calculated proton rapidity spectrum reproduces simultaneously
values observed in the NA44 experiment \cite{jacak}. 
In the case of the $Pb+Pb$ data the
transverse momentum spectra were not involved in finding a parameter
set for the fit of experimental data and are therefore predictions. We
mention that unlike in the presentation of ref. \cite{bernd7}, we now account
in our calculations for the asymmetric $n/p$ ratio, which is
$125/82$ for $Pb$ nuclei and which results in an approximately 20$\%$ smaller
proton spectrum than already shown in \cite{bernd7}.
The feature of higher stopping in the $Pb+Pb$ 
collision scenario is
nicely reflected in the rapidity spectra of the protons. While the
rapidity spectrum for protons in the $S+S$ case has a minimum at
$y=0$, for $Pb+Pb$, the rapidity spectrum of protons shows a
maximum.\\  

The calculation of Bose-Einstein correlations (BEC) was
performed using the formalism outlined in refs.
\cite{bernd2,bernd3,bernd1} including the decay of resonances. The
hadron source is assumed to be fully chaotic.  Here we present results
for pion BEC only. Kaon correlations have been discussed in refs.
\cite{bernd7,bernd2,bernd3}. In order to extract effective hadron source 
radii, we fit our results to the Gaussian form\footnote{If one performs 
a fit to a BEC function in more than one dimension, eq. (\ref{eq:fit}) 
does not represent the most general expression, because of the existence 
of an ``out-longitudinal'' cross term \cite{chapman}.}
which has been widely used by experimentalists for the
presentation of their BEC data (for the choice of the variables, {\it
cf., e.g.} ref. \cite{bernd3}):
\begin{equation}
C_2(\vec{k}_1,\vec{k}_2)\:=\: 1\:+\:\lambda\:\exp
\left[-\frac{ 1}{2} (q_\parallel^2 R_\parallel^2 \:+\:q_{side}^2
R_{side}^2\:+\:q_{out}^2 R_{out}^2) \right] \:.
\label{eq:fit}
\end{equation} 

It should be emphasized that in the present model $\lambda$ does {\it
not} represent the effect of coherence, but the momentum-dependent
effective reduction of the intercept due to the contributions from
the decay of long-lived resonances ({\it cf.}
\cite{bernd7},\cite{bernd2}).\\

Figs. 5a and 5b show our calculations for the effective radii
$R_\parallel$, $R_{side}$ and $R_{out}$ as functions of rapidity $y_K$
and transverse momentum $K_\perp$ of the pion pair compared to the
corresponding NA35 and preliminary NA49 data \cite{alber,QM95}, 
respectively. In order to make possible a comparison of the calculated 
effective radii with the experimentally obtained ones, we have to account 
for detector acceptances. In the case of $S+S$ at 200 $AGeV$ the effective 
radii as a function of $y_K$ have been calculated at $K_\perp=200 MeV$, 
while the effective radii as a function of $K_\perp$ have been calculated at
$y_K=4.0-y_{cm} \approx 1.0$ . In the case of $Pb+Pb$ at 160 $AGeV$ the
effective radii as a function of
$K_\perp$ were calculated at $y_K=4.5-y_{cm} \approx 1.6$. The
effective longitudinal radii $R_\parallel$ as a function of $K_\perp$
are evaluated in the longitudinal comoving system (LCMS).
All our
calculations, which in the case of $S+S$ have been true predictions,
 agree surprisingly well with the data. In refs.
\cite{bernd7,bernd2,bernd3} we have shown also the effective radii
for both types of heavy-ion collisions for $y_K=0.0$ and $K_\perp=0.0$. 
There the effective radii take
even larger values compared to the ones we show here when comparing
them to the data, because the contributions from resonance decays to BEC 
take their maximum values at low momenta. 
Therefore, we have strong evidence that one cannot
extract the maximal effective radii of the hadron sources from the
Bose-Einstein correlation data obtained at the CERN/SPS
since the acceptances of the experiments do not obtain data at $y_K=0.0$ 
and $K_\perp=0.0$. It is important to note that in
the case of the pion interferometry, the presence of a resonance halo
increases the size of the fireball in the central region by factors
$\sim 2.1$ ($\sim 2.0$) in  longitudinal and $\sim 1.4$ ($\sim 1.3$) in
transverse direction for
$Pb+Pb$ ($S+S$).  Values for the maximal possible resonance halo-size
are given in table 1.\\

In Fig. 6 we show a two-particle Bose-Einstein correlation function
$\tilde{C}_2(Q_{inv})$ as a function of the invariant variable, 
$Q_{inv}$. This BEC function is defined as follows ({\it cf.,} also
refs. \cite{bernd4,bernd5,grassberger}). 
Let $\rho_2(\vec{k}_1,\vec{k}_2)$ be the inclusive two-particle 
spectrum for two identical pions with momenta
$\vec{k}_1$ and $\vec{k}_2$ and $\rho_1(\vec{k})$ the inclusive
single-particle spectrum for a pion with momentum $\vec{k}$,
respectively:
\begin{equation}
\rho_1(\vec{k})\:=\:\frac{1}{\sigma}\frac{d \sigma}{d \omega}\:,\:
\rho_2(\vec{k}_1,\vec{k}_2)\:=\:\frac{1}{\sigma}\frac{d^2 \sigma} {d
\omega_1 d \omega_2}\:,\:d \omega\:=\:\frac{d^3 k}{(2\pi)^3 \cdot
2E}\:.
\label{eq:rhos}
\end{equation}

The two-particle BEC for a single pair of two pions with momenta
$\vec{k}_1$ and $\vec{k}_2$ takes the form
\begin{equation}
C_2(\vec{k}_1,\vec{k}_2))\:=\:\frac{\rho_2(\vec{k}_1,\vec{k}_2)}
{\rho_1(\vec{k}_1) \cdot \rho_1(\vec{k}_2)}
\:=\:1\:+\:\frac{\bar{c}(\vec{k}_1,\vec{k}_2)} {\rho_1(\vec{k}_1)
\cdot \rho_1(\vec{k}_2)}\:.
\label{eq:c2single}
\end{equation}

The separate contributions $\bar{c}(\vec{k}_1,\vec{k}_2)$,
$\rho_1(\vec{k}_1)$ and $\rho_1(\vec{k}_2)$ have to be calculated for
each pair of pion momenta including also the interference
contributions from the pions which originate from the decay of
resonances ({\it cf.} refs. \cite{bernd2,bernd3,bernd5}).\\

The two-particle Bose-Einstein correlation function
$\tilde{C}_2(Q_{inv})$ as a function of the invariant variable, 
$Q_{inv}$, is given by
\begin{equation}
\tilde{C}_2(Q_{inv})\:=\:1\:+\:\frac{I_2(Q_{inv})}{I_{11}(Q_{inv})}\:.
\label{eq:c2all}
\end{equation}

With $q^\mu=k_1^\mu-k_2^\mu$ we have
\begin{eqnarray} I_{11}(Q_{inv})&=&\int d\omega_1 \int
d\omega_2\:\delta
\left[ Q_{inv}\:-\:\sqrt{-q^\mu q_\mu}\: \right]
\:\rho_1(\vec{k}_1)\:\rho_1(\vec{k}_2)\:,
\nonumber\\ I_2(Q_{inv})&=&\int d\omega_1 \int d\omega_2\:\delta
\left[ Q_{inv}\:-\:\sqrt{-q^\mu q_\mu}\: \right]
\:\bar{c}(\vec{k}_1,\vec{k}_2)\:,
\label{eq:i11i2}
\end{eqnarray}

and $k^\mu$ represents the 4-momentum of a pion. The integrations
have to take into account the particular detector
acceptance for the experiment
under consideration. Here we considered the phase space 
$1.1 \leq y \leq 2.1$ and $50 MeV \leq k_\perp \leq 600 MeV$,
which covers the detector acceptance of the NA49 experiment.\\

In the previous section we mentioned that in the case of pion
interferometry we have to deal with a large fraction -- $40\%$ to
$50\%$ -- of the pions originating from resonances. 
This statement refers to an average over all particle momenta.
In case of the NA49 detector acceptance the main contributions to the
two-pion BEC come from thermal $\pi^-$ ($72.8\%$) and $\pi^-$
contributions from $\rho$- ($15.6\%$), $\omega$- ($8.7\%$) and
$\eta$-decays ($2.9\%$).
The effects of resonance decays on the two-particle Bose-Einstein 
correlation function $\tilde{C}_2(Q_{inv})$ are also shown in Fig. 6, 
where we successively have added contributions from the specific resonance
decay channels to the thermal (direct) pion contributions. The more
resonances we take into account, the narrower the correlation
function becomes because the addition of a resonance halo increases
the effective source size ({\it cf.} refs. \cite{bernd2,bernd3,csorgo}). 
As discussed in our earlier papers, the resonance contributions from the 
$\eta$-resonance yield an apparent intercept reduction. The intercept,
$I_0 = \tilde{C}_2(Q_{inv}=0)$, of the BEC takes the value 
$I_0 \approx 1 + 0.97^2 \approx 1.94$ ({\it cf.} also ref. \cite{bernd3}). 
Here we do not compare our numerical result to the preliminary data
of the NA49 Collaboration \cite{QM95}, because the current experimental 
correlation function represents a data sample of only 529 events.
Furthermore, we believe that there may be problems in the preliminary data
with the experimental two-track resolution and/or Coulomb
overcorrections that are apparent at very small values of $Q_{inv}$. 
Nevertheless the present particular calculation 
demonstrates the complexity of BEC functions as they emerge
from measurements.\\

In general, we expect the observed two-particle BEC to be strongly
dependent on detector acceptances as well on the particular
contributions from the decay of resonances. Bose-Einstein
correlations are a very complicated observable defined through
quantum statistics.  These functions depend per definition on the
choice of momenta under consideration. In general, different detector
acceptances in different experimental setups should lead for the same
heavy-ion collision at a fixed reaction energy to different results.
Thus, the interpretation of BEC measurements is also complicated. The
interpretation of extracted inverse widths (effective radii) of 
experimental BEC
depends on the specific detector acceptance under consideration.
Therefore, rather than interpret experimental fitted quantities such as
effective source radii, we propose the reader should pay attention to the
model which analyzes the data. From the hydrodynamical treatment we
learn that the hadron source (the real fireball) is represented
through a very complex freeze-out hypersurface ({\it cf.} Figs. 3a,3b). The
longitudinal and transverse extensions of the fireball change
dynamically as a function of time, rather than show up in static
effective radii. A future study \cite{future2} of the BEC for both pions
and kaons for two different detector acceptances
(which could be done by considering simultaneously the experimental
results of the NA49 and the NA44 Collaborations) is in preparation.\\

\section{Summary}

We have shown that data from two different heavy-ion experiments for
single and double inclusive cross sections of mesons and baryons can
be reproduced with a self-consistent three-dimensional relativistic
hydrodynamic description assuming an equation of state with a
phase transition to a QGP. Our data analysis indicates a stronger
stopping and an enhanced transverse flow in the case of $Pb+Pb$
collisions compared to $S+S$ collisions at CERN/SPS energies.
In particular, the preliminary $Pb+Pb$ data can be explained by
simple scaling assumptions in the initial conditions coming from
$S+S$, although the final distributions do not show these scaling
features ({\it e.g.}, compare the final rapidity spectra of protons 
({\it cf.} Fig. 4a)). 

Bose-Einstein correlation functions for pions have also been calculated.  
The NA35 and NA49 data on interferometry are surprisingly well described. 
Because the largest contributions to BEC from resonance decays are at small
particle momenta, the current BEC experiments at CERN do not measure the 
maximal possible interferometry radii for identical pions.
We also exhibit the freeze-out 
hypersurfaces; these surfaces rather than the effective
radii extracted from experimentally measured BEC functions represent 
the true space-time geometries of the sources. This caveat
should always be kept in mind when trying to interpret the BEC data.

The results of this work constitute further evidence that heavy-ion
collisions in the SPS region show fluid dynamical behaviour and can
be described by assuming an equation of state with a phase transition 
from QGP to hadronic matter. This result has to be regarded in line with 
previous hydrodynamical studies of the Marburg group
\cite{bernd7,udo,jan,bernd2,axel,udo91} and from other groups
\cite{many}. Of course, the EOS we have chosen might not be the
only one which is able to describe the current SPS heavy-ion data. 
Further analysis which allows for different equations of state
is in preparation \cite{future3}.\\[3ex]

We would like to thank B.V. Jacak, J.P. Sullivan and N. Xu for many
helpful discussions. This work was supported by the University of
California and the Deutsche Forschungsgemeinschaft (DFG). 
B.R. Schlei acknowledges a DFG postdoctoral fellowship and R.W.
is indebted to A. Capella for the hospitality extended to him at the 
LPTHE, Univ. Paris-Sud.

\newpage
\noindent
{\Large {\bf Figure Captions}}\\

\begin{description}

\item[Fig. 1] Equation of state with phase-transition to a
quark-gluon plasma. Plotted are the speed of sound, $c_0^2$, and the energy
density, $\epsilon$, as a function of temperature, $T$ \cite{dipljan}.
The curves are a result of a fit \cite{phdudo} to lattice QCD 
calculations \cite{redlich}.

\item[Fig. 2a] Initial distributions of energy and
baryon density, as well as the rapidity, normalized to their maximum
values ({\it cf.} table 1)and plotted against the longitudinal 
coordinate $z$.

\item[Fig. 2b] Boundaries in the two-parameter planes for the relative
fraction $K_L$ of thermal energy in the central fireball and the
rapidity $y_m$ at the maximum of the initial baryon distribution,
respectively.
The crossed lines indicate our particular choices for the calculations.

\item[Fig. 3a] Time contour plots of the freeze-out hypersurfaces in
the $z-r$ plane.
Each line represents the freeze-out hypersurface at a fixed time $t$.

\item[Fig. 3b] Three-dimensional view of the freeze-out hypersurfaces.

\item[Fig. 3c] Three-dimensional view of the transverse velocity fields
at freeze-out.

\item[Fig. 4a] Rapidity spectra for negative hadrons ($h^{-}$) and protons
($p$). The data points of the NA35 data are taken from \cite{wenig}. 
The data points of the preliminary results of the NA49
Collaboration for negative hadrons from central $Pb+Pb$ collisions at 160 $%
AGeV$ are shown with two different markers.
Circles (diamonds) stand for the measurements from the VTP2 (MTPC)
({\it cf.} refs. \cite{NA49_1,NA49_2}).

\item[Fig. 4b] Transverse momentum spectra of negative hadrons ($h^{-}$)
and protons ($p$). The data points of the NA35 data are taken from 
\cite{wenig}.
For $Pb+Pb$ at 160 $AGeV$ the integrations with respect to rapidity $y$
have been performed over the interval $|y|\leq 3.0$.

\item[Fig. 5a]  Effective radii extracted from Bose-Einstein correlation
functions as a function of the rapidity $y_K$ of the pair and the
transverse average momentum $K_{\perp }$ of
the pair for all pions compared to NA35 data \cite{alber,QM95}.

\item[Fig. 5b] Effective radii extracted from Bose-Einstein correlation
functions as a function of the transverse average momentum $K_{\perp }$ of
the pair for all pions compared to preliminary NA49 data \cite{QM95}.

\item[Fig. 6] Two-particle Bose-Einstein correlation functions
$\tilde{C}_2(Q_{inv})$ of negatively charged pions. The contributions from
resonance decays are successively added to the correlation function of
thermal $\pi ^{-}$ (dotted lines). The resultant correlation function 
of all $\pi ^{-}$ is given by the solid line. The integrations have been 
performed with respect to the detector acceptance of the NA49 experiment
(see text).

\end{description}

\newpage
\vspace{1.5cm}

\noindent
{\Large {\bf Table Caption}}
\begin{description}
\item[Table 1]  Properties of initial fireball extracted from a
hydrodynamical analysis of
the $S+S$ NA35 data \cite{wenig} and the $Pb+Pb$ NA49
data \cite{NA49_1,NA49_2}.
\item[Table 2]  Chemical compositions of mesons, baryons and anti-baryons
at freeze-out for $S+S$ at 200 $AGeV$ and $Pb+Pb$ at 160 $AGeV$.
Listed are their relative fractions.
\end{description}

\newpage
\vspace{2.5cm}
\centerline{\huge \bf Table 1}
\label{tab1}

\begin{table}
\begin{center}
\begin{tabular}{|l|l|l|}
\hline
 & S+S & Pb+Pb \\
\hline
\multicolumn{3}{|c|}{\bf Fit parameters}\\
\hline
Rel. fraction, $K_L$, of thermal energy in the central fireball 
& 0.43 & 0.65 \\
\hline
Longitudinal extension, $\Delta$, of central fireball & 0.6 $fm$ & 1.2 $fm$ \\
\hline
Rapidity, $y_\Delta$, at edge of central fireball & 0.9 & 0.9 \\
\hline
Rapidity, $y_m$, at maximum of initial baryon distribution & 0.82 & 0.60 \\
\hline
Width, $\sigma$, of initial baryon $y$-distribution & 0.4 & 0.4 \\
\hline
\multicolumn{3}{|c|}{\bf Output}\\
\hline
Center of mass rapidity, $y_{cm}$ & 3.03 & 2.92 \\
\hline
Max. initial energy density, $\epsilon_{max}$ 
& 13.0 $GeV/fm^3$ & 20.4 $GeV/fm^3$ \\
\hline
Max. initial baryon density, $B^0_{max}$ 
& 2.66 $fm^{-3}$ & 3.20 $fm^{-3}$ \\
\hline
Rel. fraction of baryons in central fireball & 0.49 & 0.73\\
\hline
Max. lifetime & 6.9 $fm/c$ & 13.5 $fm/c$ \\
\hline
Initial volume of QGP & 24 $fm^3$ & 174 $fm^3$ \\
\hline
Lifetime of QGP & 1.5 $fm/c$ & 3.4 $fm/c$ \\
\hline
Max. halo-size in ``longitudinal'' direction & 2.5 $fm$ & 5.6 $fm$ \\
\hline
Max. halo-size in ``side'' direction  & 1.0 $fm$ & 2.4 $fm$ \\
\hline
Max. halo-size in ``out'' direction & 1.3 $fm$ & 2.6 $fm$ \\
\hline
\end{tabular}\\
\end{center}
\end{table}

\newpage

\begin{table}
\begin{center}
\begin{tabular}{|l|c|r|r|}
\hline
Particle species & Mass [$GeV/c^2$] & S+S & Pb+Pb \\
\hline
\hline
$\pi$ (stable)       & 0.139 & 51.094 \% & 49.201 \% \\
\hline
$\omega$             & 0.783 &  2.452 \% &  2.304 \% \\
\hline
$\eta^\prime$        & 0.958 &  0.297 \% &  0.279 \% \\
\hline
$\eta$               & 0.549 &  2.941 \% &  2.765 \% \\
\hline
$\rho$               & 0.770 &  7.429 \% &  6.980 \% \\
\hline
$K^+$ (stable)       & 0.494 &  7.689 \% &  7.763 \% \\
\hline
$K^0$                & 0.498 &  2.573 \% &  2.582 \% \\
\hline
$\Lambda$            & 1.116 &  1.174 \% &  1.356 \% \\
\hline
$\Sigma$             & 1.193 &  2.232 \% &  2.578 \% \\
\hline
$\Delta$             & 1.232 &  4.115 \% &  5.014 \% \\
\hline
$N$ (stable)         & 0.939 &  9.009 \% & 10.960 \% \\
\hline
$K^\star$            & 0.893 &  0.952 \% &  0.954 \% \\
\hline
$\Sigma^\star$       & 1.386 &  0.613 \% &  0.707 \% \\
\hline
$\Xi$                & 1.320 &  0.497 \% &  0.541 \% \\
\hline
$K^-$ (stable)       & 0.494 &  4.400 \% &  3.881 \% \\
\hline
anti-$K^0$           & 0.498 &  1.472 \% &  1.289 \% \\
\hline
anti-$\Lambda$       & 1.116 &  0.049 \% &  0.036 \% \\
\hline
anti-$\Sigma$        & 1.193 &  0.092 \% &  0.068 \% \\
\hline
anti-$\Delta$        & 1.232 &  0.098 \% &  0.066 \% \\
\hline
anti-$N$ (stable)    & 0.939 &  0.215 \% &  0.147 \% \\
\hline
anti-$K^\star$       & 0.893 &  0.546 \% &  0.479 \% \\
\hline
anti-$\Sigma^\star$  & 1.386 &  0.025 \% &  0.019 \% \\
\hline
anti-$\Xi$           & 1.320 &  0.036 \% &  0.028 \% \\
\hline
\end{tabular}\\
\end{center}
\vspace{2.5cm}
\centerline{\huge \bf Table 2}
\label{tab2}
\end{table}
\end{document}